\documentclass[sigconf]{acmart}

\usepackage{booktabs}
\usepackage{enumitem}
\usepackage{graphicx}
\usepackage{tabularx}
\usepackage[most]{tcolorbox}

\newtcolorbox{agentpromptbox}[1][]{
  enhanced,
  breakable,
  colback=blue!2,
  colframe=blue!45!black,
  boxrule=0.6pt,
  arc=1.5mm,
  left=2mm,
  right=2mm,
  top=1.5mm,
  bottom=1.5mm,
  fonttitle=\bfseries,
  title={#1}
}

\AtBeginDocument{%
  }

\setcopyright{none}
\renewcommand\footnotetextcopyrightpermission[1]{}

\acmISBN{}
\acmDOI{}

\begin{document}

\title{Are Production Cloud Skills Adequately Tested? Measuring and Governing Skill Test Adequacy in Practice}

\author{Haotian Si}
\affiliation{%
  \institution{Alibaba Cloud}
  \city{Hangzhou}
  \country{China}
}

\author{Junyi Chen}
\affiliation{%
  \institution{Alibaba Cloud}
  \city{Hangzhou}
  \country{China}
}

\author{Shuyang Yu}
\additionalaffiliation{%
  \institution{Columbia University}
  \city{New York}
  \country{USA}
}
\affiliation{%
  \institution{Alibaba Cloud}
  \city{Hangzhou}
  \country{China}
}

\author{Ruifeng Nie}
\additionalaffiliation{%
  \institution{Aalto University}
  \city{Espoo}
  \country{Finland}
}
\affiliation{%
  \institution{Alibaba Cloud}
  \city{Hangzhou}
  \country{China}
}

\author{Jiate Li}
\affiliation{%
  \institution{Alibaba Cloud}
  \city{Hangzhou}
  \country{China}
}

\author{Jianqiang Zhao}
\affiliation{%
  \institution{Alibaba Cloud}
  \city{Hangzhou}
  \country{China}
}

\author{Meng Li}
\affiliation{%
  \institution{Alibaba Cloud}
  \city{Hangzhou}
  \country{China}
}

\author{Dengcheng He}
\affiliation{%
  \institution{Alibaba Cloud}
  \city{Hangzhou}
  \country{China}
}

\renewcommand{\shortauthors}{Si et al.}
\begin{abstract}
Cloud platforms increasingly deliver reusable Cloud Skills that guide AI agents through multi-step resource operations, user choices, validation, and recovery. Existing Skill evaluation primarily measures whether a Skill improves task success, but passing the available testcases does not reveal which behaviors specified by the Skill remain untested. We introduce \emph{Skill Test Adequacy}, a scenario-conditioned criterion that evaluates a test suite against the complete set of operational test obligations specified by a Skill. Given a Skill package and normalized testcases containing a prompt, an initial resource state, and expected user decisions, the assessment determines whether each obligation is exercised by at least one testcase scenario; the resulting records provide both a suite-level score and explicit test gaps. We operationalize the criterion through parallel obligation proposals, disagreement-preserving aggregation, testcase-level status proposals, expert review, and source-grounded recommendations. Alibaba Cloud deploys this process as a mandatory gate before task-success evaluation and subsequent release checks. Among 157 initial assessments recorded before gate-driven remediation, 57 (36.3\%) fall below the mandatory 80\% gate and 76 (48.4\%) remain below the recommended 90\% level. The process also produces 132 reports containing 639 obligation-level recommendations, with a median of four per Skill. Finally, we release \emph{SkillAdeqBench}, an exploratory subset of the reviewed records for studying automatic adequacy assessment. Skill Test Adequacy complements task-success evaluation by making the untested scope of production Cloud Skills explicit.
\end{abstract}



\maketitle

\section{Introduction}

Cloud platforms are moving toward an agentic operating model. Instead of requiring users to manually navigate service consoles, API documentation, command-line tools, and troubleshooting guides, large language model agents can help users create cloud resources, configure permissions, deploy applications, set alerts, and diagnose failures. Reliable operation requires more than a capable base model: agents also need accurate, service-specific knowledge about workflows, dependencies, constraints, and best practices.

\begin{figure*}[!t]
  \centering
  \includegraphics[width=\textwidth]{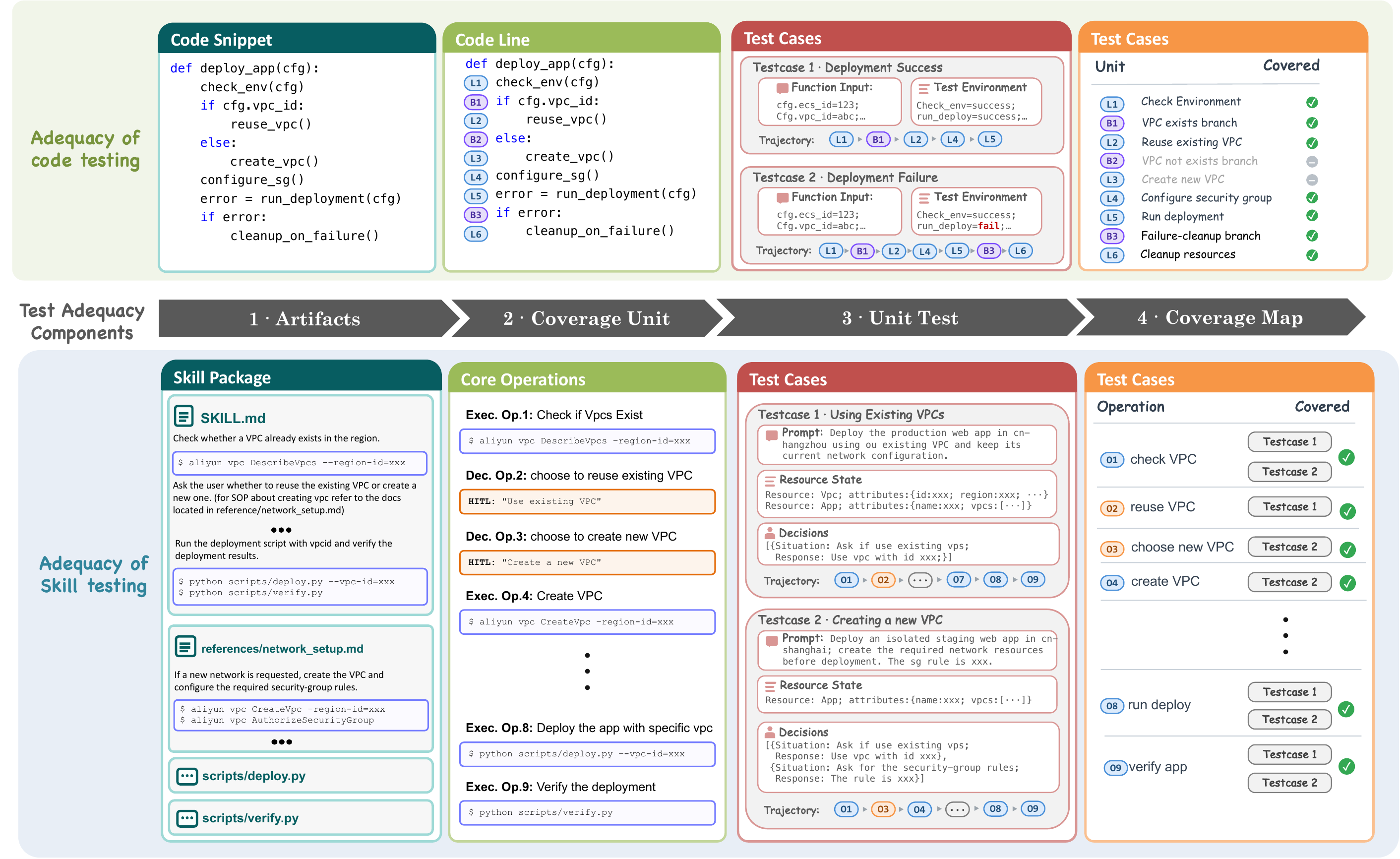}
  \caption{Code Coverage and Skill Test Adequacy. Code exposes executable lines and branch outcomes as coverage units and supports runtime instrumentation. A Cloud Skill expresses workflow-relevant operations through natural-language instructions, commands, and user-mediated choices distributed across \texttt{SKILL.md}, reference files, and scripts. Different testcase prompts, initial resource states, and expected user decisions select different specified operation paths; suite-level adequacy aggregates the covered operations across all testcase scenarios.}
  \label{fig:coverage-comparison}
\end{figure*}

Cloud platforms provide this knowledge through reusable \emph{Cloud Skills for AI agents}\footnote{See the Alibaba Cloud Agent Skills portal at \url{https://skills.aliyun.com/}.}, which we refer to as \emph{Cloud Skills} for brevity. A recent empirical study of 238 real-world Skills identified stepwise instructions as a common component: among the 224 \texttt{SKILL.md} files included in its H2-level taxonomy analysis, 49.2\% contained instructions that decompose a workflow into manageable steps~\cite{hong2026anatomy}. This workflow-oriented form is particularly relevant to cloud operations, where completing a task commonly requires multiple stateful resource operations, user decisions, validation steps, and failure handling.

Once released through a cloud platform, a Skill becomes a production delivery artifact. It is reviewed, tested, versioned, maintained, and relied upon as part of a user-facing capability. Incorrect guidance can cause unsafe configuration, unnecessary cost, incomplete recovery, or changes to the wrong resource. This lifecycle changes the quality question. Experimental evaluation asks whether providing a Skill improves an agent's task performance. Production release must additionally ask whether the Skill itself has been tested comprehensively.

Consider a Skill for deploying a web application on Elastic Compute Service (ECS). It may instruct the agent to check whether a virtual private cloud (VPC) exists, create one when needed, configure a security group, create an ECS instance, bind a public IP address, validate deployment, and clean up resources after partial failure. A testcase scenario whose initial state already contains a suitable VPC and security group may cover a successful deployment path. It does not cover VPC creation, security-group creation, or failure cleanup. Passing the selected testcase therefore does not reveal which behaviors specified by the Skill have never been tested.

Conventional software testing makes such omissions visible through test-adequacy criteria. Source code provides executable lines and branch outcomes; behavioral models provide states and transitions; API specifications provide operations and parameters. Workflow-oriented Cloud Skills, however, are multi-file natural-language packages. They provide neither explicit test obligations nor explicit links from testcases to the behavior they exercise. Workflow instructions may also coexist with API references, examples, scripts, parameter tables, and troubleshooting notes, while the status of an obligation depends on the testcase's initial resource state and expected user decisions.

Figure~\ref{fig:coverage-comparison} contrasts this problem with code coverage. In source code, executable syntax exposes lines and branch outcomes before testing, and instrumentation records which units execute. A Skill package instead distributes operational behavior across its main workflow description, supporting reference files, and executable scripts. A reference file may elaborate concrete workflow operations and therefore contribute test obligations rather than merely provide background material. The resulting inventory includes both executable commands and user-mediated alternatives; distinct choices that induce different behavior are represented as separate obligations. An adequacy assessment must identify these implicit obligations and then evaluate each one for a scenario jointly specified by a prompt, an initial resource state, and expected user decisions. Direct code instrumentation therefore does not transfer to this setting.

Table~\ref{tab:coverage-comparison} isolates the resulting measurement challenges. They concern not only how per-obligation status is determined, but also how a valid denominator and a reviewable scenario-to-obligation judgment are established.

We address this missing engineering capability by introducing \emph{Skill Test Adequacy}. The operational behaviors specified by a Skill form its test obligations. For each obligation, an adequacy assessment determines whether the testcase scenario defined by its prompt, initial resource state, and expected user decisions is expected to exercise that behavior under the Skill specification. The fraction of obligations covered by at least one testcase provides a suite-level adequacy score, while the obligation-level records expose the specific behaviors that remain untested. Adequacy does not prove correctness; it makes the scope of existing tests inspectable.

To apply the criterion in production, we instantiate it as a specification-driven audit process. Independent agents propose obligation inventories under the same adequacy standard, and aggregation makes agreements and disagreements available for review. After reviewers establish the final obligations, an agent proposes a complete covered/uncovered assignment for each testcase from the Skill and its scenario specification. Reviewers correct these assignments, and only the resulting records determine the reported adequacy. Confirmed gaps are converted into source-grounded actions for adding or revising tests, clarifying the Skill, or recording an explicit exception.

\begin{table*}[!t]
\centering
\caption{The principal measurement challenges differ between code coverage and Skill Test Adequacy.}
\label{tab:coverage-comparison}
\small
\begin{tabularx}{\textwidth}{p{0.20\textwidth}XX}
\toprule
\textbf{Measurement challenge} & \textbf{Code coverage} & \textbf{Skill Test Adequacy} \\
\midrule
Denominator validity &
The selected syntactic criterion determines the denominator mechanically. &
The denominator is a semantic interpretation: omitting a low-salience obligation creates false confidence, while admitting supporting material creates false gaps. \\
\midrule
Available evidence &
Instrumentation observes events from an actual execution. &
The measurement must determine which behavior a testcase scenario requires under the Skill without treating its declarative scenario fields as a runtime trace. \\
\midrule
Latent control context &
Branch sites and outcomes have explicit identities in the control-flow graph. &
Resource conditions and user choices may encode alternatives only in prose; identical commands can therefore denote different obligations. \\
\midrule
Uncertainty and review &
Given correct instrumentation, unit execution is normally a deterministic observation. &
Natural-language ambiguity can affect both obligation recovery and scenario mapping, requiring source-grounded disagreement handling and review. \\
\midrule
Evolution and audit &
Stable program locations support mapping execution records across revisions. &
Prose edits, cross-file relocation, and changed obligation granularity require semantic normalization and retained provenance. \\
\bottomrule
\end{tabularx}
\end{table*}

The reviewed records also provide material for studying how this assessment can be automated. After reporting the industrial findings, Section~\ref{sec:benchmark} describes the exploratory benchmark derived from a subset of these records.

This paper makes three contributions:
\begin{itemize}[leftmargin=*]
  \item We define \emph{Skill Test Adequacy}, including the operational test obligations derived from workflow-oriented Cloud Skills, their scenario-conditioned covered/uncovered status, the suite-level adequacy score, and the claim boundary.
  \item We operationalize the criterion as an auditable release practice that combines parallel obligation recovery, disagreement-preserving aggregation, scenario-conditioned status proposals, and source-grounded review to establish testcase-level adequacy records and remediation worklists.
  \item We report deployment evidence from 157 Cloud Skills under development since the audit process went online, including their initial adequacy assessments, release-gate intervention, and the obligation-level recommendation workload produced by the platform.
\end{itemize}

\section{Skill Test Adequacy}

\subsection{Scope and Semantic Domain}

We focus on \emph{workflow-oriented Cloud Skills}: packages that specify how an AI agent should complete a user-facing cloud task through observable cloud operations, user-mediated choices, validation, or recovery. A workflow may span multiple files, and scripts or reference documents may supply executable details. We exclude artifacts that only enumerate API syntax, provide background knowledge, or answer open-ended questions without specifying operational behavior.

Let $\mathcal{X}$ be the set of cloud and interaction states and $\mathcal{A}$ the set of observable operations. A conforming execution of Skill $S$ is a finite trace
\[
  \sigma=x_0\,a_1\,x_1\cdots a_m\,x_m,
  \qquad x_i\in\mathcal{X},\;a_i\in\mathcal{A}.
\]
The semantic language $\mathcal{L}(S)$ contains all traces permitted by the Skill's workflow instructions. This language is a specification-level object: it describes how a correct agent may execute the Skill, rather than recording what one runtime agent actually did.

Each normalized testcase $t=(p,r,d)$ combines a user prompt $p$, an initial resource state $r$, and expected user decisions $d$ supplied at interactive points in the workflow. We write
\[
  \Gamma(t)=\mathcal{F}(p)\cup r\cup\mathcal{D}(d)
\]
for its scenario facts and constrain the Skill semantics to
\[
  \mathcal{L}_S(t)=
  \{\sigma\in\mathcal{L}(S)\mid x_0\models r
  \land \sigma\models\mathcal{F}(p)\cup\mathcal{D}(d)\}.
\]
Thus, $\mathcal{L}_S(t)$ contains exactly the conforming Skill executions compatible with the requested task, initial cloud state, and user responses that determine interactive branches.

\subsection{Operational Test Obligations}

Let $\mathcal{B}(S)$ contain the behavior clauses semantically entailed by the Skill package. A clause is release-relevant when it (i) participates in a user-facing workflow, (ii) prescribes behavior rather than merely describing an interface, (iii) has an identifiable activation condition, and (iv) denotes an observable and testable effect. An unconditional behavior uses the activation predicate $\mathsf{true}$. We write the release-relevant semantic domain as
\[
\begin{aligned}
  \mathcal{R}(S)=\{b\in\mathcal{B}(S)\mid{}\;&
  \mathsf{workflow}(b)\land\mathsf{prescriptive}(b)\\
  &{}\land\mathsf{triggerable}(b)\land\mathsf{observable}(b)\\
  &{}\land\mathsf{testable}(b)\}.
\end{aligned}
\]
This criterion excludes detached API syntax, background explanation, and environment-wide setup or teardown that does not belong to the user-facing workflow.

An operational test obligation is the normalized record
\[
  o=\langle \alpha_o,g_o,q_o,\ell_o\rangle.
\]
Here, $\alpha_o$ identifies an observable operation or user interaction; $g_o$ is an activation predicate over the testcase facts and the execution prefix, capturing resource conditions, user choices, failure states, and workflow context; $q_o$ is the expected result predicate; and $\ell_o$ is the nonempty set of source spans supporting the obligation. The obligation universe is
\[
  O(S)=\mathcal{R}(S)/{\equiv},
\]
where two clauses denote the same obligation exactly when
\[
  b\equiv b'
  \;\Leftrightarrow\;
  \alpha_b\simeq\alpha_{b'}
  \land (g_b\leftrightarrow g_{b'})
  \land (q_b\leftrightarrow q_{b'}).
\]
Thus, paraphrases of the same guarded behavior collapse into one obligation, whereas identical commands remain distinct when they occur under different guards or establish different results.

A valid obligation inventory must discharge two specification-level conditions. \emph{Soundness} requires every normalized obligation to be grounded in at least one release-relevant Skill clause:
\[
  \forall o\in O(S):
  \ell_o=
  \{\mathsf{source}(b)\mid b\in\mathcal{R}(S),\,b\equiv o\}
  \neq\varnothing.
\]
\emph{Completeness and non-redundancy} require every release-relevant clause to map to exactly one normalized obligation:
\[
  \forall b\in\mathcal{R}(S),\ \exists!\,o\in O(S): b\equiv o.
\]
These conditions make the denominator a property of the Skill semantics rather than of a particular extraction algorithm.

For a trace $\sigma$ and testcase $t$, obligation realization is defined by
\[
\begin{split}
  \mathsf{match}(o,\sigma,t)\;\Leftrightarrow\;
  \exists i:\;&a_i\simeq\alpha_o \;\land\\
  &g_o(\sigma_{<i},\Gamma(t))\;\land\;q_o(x_i),
\end{split}
\]
where $\simeq$ denotes semantic operation identity rather than string equality.

Operational obligations recur as resource operations, user-mediated operations, validation operations, and recovery operations. These forms support interpretation but do not define separate metrics. The compact review record represents $\alpha_o$, its distinguishing context, and $q_o$ through \texttt{command}, \texttt{label}, and \texttt{result}; provenance $\ell_o$ is retained by the audit platform.

\subsection{Scenario-Conditioned Obligation Status}

The binary status of an obligation is defined only for a well-formed testcase. Its facts must be consistent and the scenario must admit at least one conforming execution:
\[
  \mathsf{Sat}_S(t)
  \;\Leftrightarrow\;
  \Gamma(t)\not\models\bot
  \land \mathcal{L}_S(t)\neq\varnothing.
\]
It must also be \emph{decision-complete} for the obligation inventory:
\[
  \forall o\in O(S):
  \left|
  \{\mathsf{match}(o,\sigma,t)\mid\sigma\in\mathcal{L}_S(t)\}
  \right|=1.
\]
This condition permits inconsequential variation in ordering or tool choice, but rejects a testcase whose missing resource condition or user choice leaves an obligation covered on some conforming executions and uncovered on others.

For testcases satisfying $\mathsf{Sat}_S(t)$ and decision completeness, we define
\[
  \mathsf{cov}_S(t,o)=1
  \;\Leftrightarrow\;
  \forall\sigma\in\mathcal{L}_S(t):
  \mathsf{match}(o,\sigma,t).
\]
Equivalently, each matrix entry checks the specification-level verification condition
\[
  S,\Gamma(t)\models\Diamond o,
\]
where $\Diamond o$ means that every conforming execution selected by the testcase eventually realizes $o$. This universal interpretation prevents a testcase from receiving credit merely because one possible agent trajectory could realize the operation.

This relation supplies the per-operation coverage status used by the adequacy assessment. In plain terms, it asks whether the testcase scenario is expected to exercise the operation under the Skill specification, rather than whether the operation happened to appear in one observed agent execution.

Let $T=\{t_1,\ldots,t_m\}$ and $O(S)=\{o_1,\ldots,o_n\}$. The testcase--obligation matrix is
\[
  M_{ij}=\mathsf{cov}_S(t_i,o_j)\in\{0,1\}.
\]
An obligation is covered by the suite when at least one testcase discharges it:
\[
  c_j=\bigvee_{i=1}^{m}M_{ij}.
\]
Skill Test Adequacy and the reviewed gap set are therefore
\[
  \mathit{STA}(S,T)=\frac{1}{n}\sum_{j=1}^{n}c_j,
  \qquad
  G(S,T)=\{o_j\in O(S)\mid c_j=0\}.
\]

\subsection{Properties and Claim Boundary}

The definition separates two verification responsibilities. Establishing $O(S)$ checks the soundness and completeness of the adequacy denominator; establishing $M_{ij}$ checks whether the testcase facts entail realization of a particular obligation. The reviewed report is valid only when both responsibilities are discharged.

The resulting adequacy score has three useful sanity properties. It is bounded, $0\leq\mathit{STA}(S,T)\leq1$; it is monotonic under test addition,
\[
  T\subseteq T'
  \Rightarrow
  \mathit{STA}(S,T)\leq\mathit{STA}(S,T');
\]
and it is complete with respect to the selected obligation criterion,
\[
  \mathit{STA}(S,T)=1
  \Leftrightarrow
  G(S,T)=\varnothing.
\]
The criterion is independent of the measurement implementation. Candidate generation and review mechanisms help establish these semantic objects but do not change their definitions.

Skill Test Adequacy is a scope-based adequacy signal, not proof of runtime correctness. A covered obligation may still be executed with incorrect parameters, checked by a weak oracle, or implemented unsafely. The score does not enumerate every parameter combination, prove arbitrary ordering properties, or validate runtime safety. These concerns require traces, assertions, formal constraints, risk analysis, or further review. Likewise, the operational 80\% release threshold is a release policy, not a theoretical sufficiency bound.

\section{Operationalizing Skill Test Adequacy}

\subsection{Overview}

\begin{figure*}[t]
  \centering
  \includegraphics[width=\textwidth]{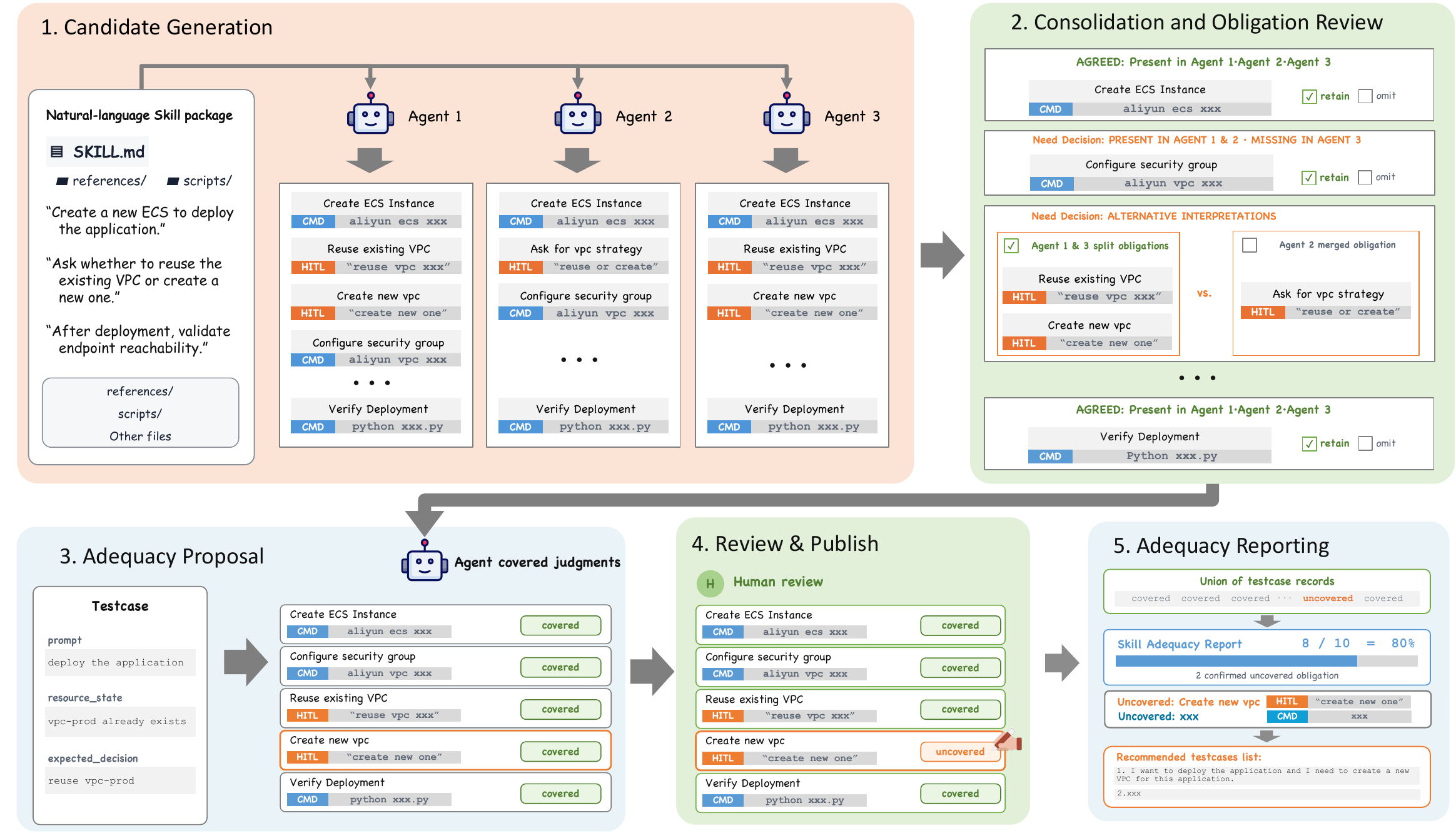}
  \caption{Operationalizing Skill Test Adequacy in the release process. Independent agents propose obligation inventories under the same adequacy standard. Aggregation normalizes agreement while preserving the union and disagreements for review. Reviewers establish the obligation set; an agent then proposes per-testcase covered/uncovered statuses for reviewer correction. Deterministic aggregation computes the suite-level adequacy score, after which confirmed gaps are converted into source-grounded test-improvement recommendations.}
  \label{fig:pipeline}
\end{figure*}

Our production process applies the criterion to a Skill package and its normalized testcases and produces four artifacts: a reviewed obligation set, one adequacy record set per testcase, a suite-level adequacy report, and source-grounded remediation recommendations. Figure~\ref{fig:pipeline} summarizes this operationalization. Agents first propose candidate obligations, aggregation makes agreement and disagreement visible, and reviewers establish the final obligation set. An agent then proposes per-testcase statuses for reviewer correction; the resulting records determine the adequacy report and recommendations.

The agent instances used for obligation candidate generation and testcase adequacy proposal are powered by \textbf{Qwen3.7-Max}. Each independent candidate run uses a fresh agent session so that agreement reflects repeated analysis of the Skill rather than shared conversational state.

\subsection{Parallel Candidate Generation}

The pipeline reads the complete Skill folder rather than only its main file. This is necessary because workflow instructions may invoke scripts or rely on reference files for command semantics. At the same time, not every operation-like phrase defines a test obligation. Candidate recovery must distinguish task workflow from supporting material.

We run three agents concurrently and independently for each Skill. Every agent receives the same adequacy standard and follows an observe--analyze--inspect loop: it reads the relevant package files, identifies the operations that participate in user-facing workflows, revisits source material when the boundary or granularity is unclear, and emits a complete candidate inventory. The standard includes cloud-resource operations, official scripts and external invocations, user choices that affect subsequent execution, result validation, and recovery or cleanup. It also requires each candidate to be independently testable and represented by a command, a workflow-distinguishing label, and an expected result.

The parallel runs are redundancy rather than votes. Their purpose is to increase candidate recall and reveal unstable interpretations before a reviewer commits to an obligation boundary. A candidate omitted by one run remains available when another run grounds it in the Skill. Failed runs are retried and do not silently shrink the candidate pool.

\subsection{Consolidation and Obligation Review}

The aggregator compares the independent inventories against the Skill package. Semantically equivalent candidates with the same workflow role are normalized into one item. The union is retained rather than reduced by majority vote, because an operation recovered by only one agent may be a genuine low-salience obligation. Incompatible granularities or alternative interpretations are grouped as disagreements and shown together. Aggregation therefore organizes review; it does not determine the reference answer.

The labeling platform shows the original Skill files beside compact obligation cards. Candidates with consistent semantics across runs are preselected, while disagreements and one-sided candidates are highlighted. A reviewer checks each item against the Skill and may add an omitted operation, delete a reference-only item, merge over-split candidates, or split an operation whose alternatives can be tested independently. Review also removes environment-wide setup and teardown that is detached from the user-facing workflow.

The accepted cards form $O(S)$. The reviewed obligation inventory determines the adequacy denominator.

\subsection{Adequacy Proposal and Review}

For each testcase, an agent receives the complete Skill package, the reviewed obligations, the testcase prompt, its initial resource state, and its expected user decisions. The prompt identifies the requested task, the resource state determines which resource-dependent behavior is applicable, and the expected decisions select user-mediated branches. For example, the presence of an appropriate VPC makes reuse applicable, while an expected response to reuse that VPC selects the reuse branch rather than the create-new branch.

The agent again follows an iterative inspection process. It first checks that the obligation inventory is complete, then analyzes the testcase scenario, revisits the Skill or testcase whenever a condition remains unclear, and finally emits a complete adequacy output in which every operation is marked \texttt{covered} or \texttt{uncovered}. Resource state is treated as an initial condition, not as evidence that an operation has already executed. An item is covered only when the combined scenario is expected to exercise that operation under the Skill specification.

Reviewers inspect the proposal together with the Skill and the complete testcase scenario. Each obligation can be switched directly between \texttt{covered} and \texttt{uncovered}; edits are saved automatically. Testcases enter this stage only after their prompt, resource-state, and expected-decision fields satisfy the input contract in Section~2.1. The corrected operation set and statuses, rather than the agent proposal, form the testcase's final adequacy output.

\subsection{Adequacy Reporting and Recommendation Generation}

The final report contains the reviewed obligation list, one adequacy output per testcase, the suite-level union of covered obligations, and $\mathit{STA}(S,T)$. Suite aggregation is deterministic: an obligation is covered if any reviewed testcase labels it covered. The reviewed gap set $G(S,T)$ is computed directly from the testcase--obligation matrix in Section~2.3. Only obligations in this set, as established by the corrected human-review records, enter recommendation generation. Agent proposals that reviewers reject or correct cannot independently produce recommendations.

For each confirmed gap, the recommendation module combines the obligation's operation, workflow-distinguishing label, expected result, retained Skill provenance, and relevant testcase context. It describes the behavior that remains untested, the resource condition or user choice needed to exercise it, and the result a new or revised testcase should verify. The module may instead identify that the testcase description is underspecified, that the Skill itself requires clarification, or that the gap should be handled through an explicit exception. Every recommendation retains a link to the reviewed obligation and its source context so that an author or reviewer can inspect the basis of the advice.

The recommendation module supplies a remediation worklist rather than autonomously changing a Skill or its tests. Skill authors and release reviewers decide which action is appropriate, and adequacy is reassessed after revised testcases are reviewed. The deployment records in this paper include the recommendations generated from initial adequacy assessments; they do not establish whether individual recommendations were accepted or caused later adequacy gains.

Each adequacy record contains \texttt{command}, \texttt{label}, \texttt{result}, and \texttt{cover\_status}. Together, the records for a testcase represent the complete operation set and its scenario-conditioned coverage statuses.

\section{Industrial Study}

\subsection{Application Background}

Alibaba Cloud integrated Skill Test Adequacy into the release process for production Cloud Skills. A Skill author submits the Skill package together with its normalized test suite. The reviewed audit process establishes the obligation inventory and testcase--obligation matrix, computes the suite-level adequacy score, and returns uncovered obligations as source-grounded recommendations. This adequacy assessment is the first mandatory quality gate: only a Skill that passes it may proceed to task-success evaluation and the remaining release checks. A published Alibaba Cloud Skill has therefore passed both the adequacy gate and the downstream evaluation process. This sequencing prevents high task success on a narrow set of scenarios from bypassing broader test-adequacy requirements.

The deployment uses two policy levels. A Skill must reach a Skill Test Adequacy score of at least 80\% to pass the release gate. A Skill below this threshold is returned for remediation and reassessment and cannot enter task-success testing in the meantime. The platform additionally recommends 90\% as a target release level. The latter preserves visibility into residual gaps but does not replace the mandatory 80\% rule.

Since this process went online, the platform has evaluated 157 Cloud Skills under development. It records an \emph{initial adequacy assessment} when each Skill reaches the gate before any gate-driven remediation. The study analyzes these 157 assessments and 132 \emph{recommendation reports} generated from reviewed adequacy gaps. The analysis is descriptive and uses only the initial adequacy scores and their associated reports. It does not compare against a predeployment cohort or claim that recommendations caused subsequent improvements.

\subsection{Analysis}

\begin{figure}[t]
  \centering
  \includegraphics[width=\columnwidth]{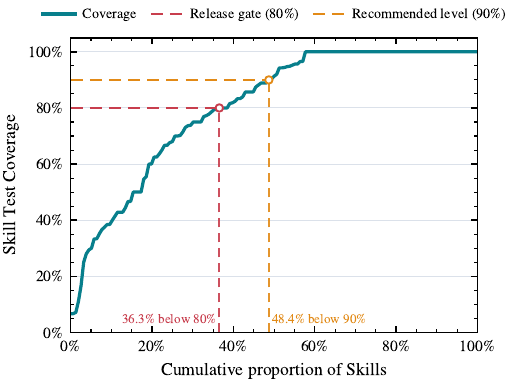}
  \caption{Ranked distribution of 157 initial Skill Test Adequacy scores. The horizontal axis is the cumulative proportion of Skills. The L-shaped guides mark the mandatory 80\% release gate and recommended 90\% level, together with the proportion of Skills below each threshold.}
  \label{fig:initial-coverage}
\end{figure}

\paragraph{Initial adequacy.}
Figure~\ref{fig:initial-coverage} shows the ranked distribution of the 157 initial adequacy scores. They have a mean of 80.3\% and a median of 91.3\%; the interquartile range is 68--100\%. Although 67 Skills (42.7\%) receive a score of 100\%, the lower tail is substantial: the 10th percentile is 39.4\%.

Only 100 of 157 Skills (63.7\%) satisfy the mandatory 80\% gate in their initial adequacy assessment; 57 (36.3\%) must be remediated before release. At the recommended 90\% level, 81 Skills (51.6\%) meet the target and 76 (48.4\%) do not. The median alone would therefore give an overly optimistic view: more than one third of Skills initially fail the release requirement, and nearly half remain below the recommended level. This lower tail demonstrates why passing sampled tasks is not an adequate proxy for the scope of a Skill's test suite.

\begin{figure}[t]
  \centering
  \includegraphics[width=\columnwidth]{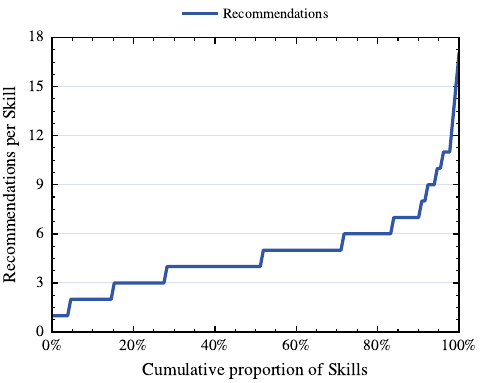}
  \caption{Ranked number of obligation-level recommendations across 132 Skills with a recommendation report. The horizontal axis is the cumulative proportion of Skills.}
  \label{fig:recommendation-workload}
\end{figure}

\paragraph{Recommendation workload.}
Across the 132 recommendation reports, the module produces 639 obligation-level recommendations. The mean is 4.84 recommendations per Skill, the median is four, and the interquartile range is three to six; the largest report contains 17 recommendations. Figure~\ref{fig:recommendation-workload} shows their ranked distribution.

The recommendation reports span all 57 Skills below the 80\% gate and 75 Skills that satisfy it. The scalar and the worklist therefore serve different release-review purposes: the threshold decides whether a Skill may proceed, while the reviewed gap set identifies what remains untested even for a passing Skill. Depending on the retained source and scenario context, the corresponding action may be to add a testcase, clarify its prompt, resource state, or expected decisions, revise an ambiguous Skill, or document an explicit exception.

These deployment records establish two claims at the current stage. First, inadequate Skill tests are common enough to affect release decisions: 36.3\% of Skills fail the gate in their initial adequacy assessment. Second, obligation-level analysis produces diagnostic information beyond the score, including for Skills that already satisfy the release threshold. The data do not yet establish recommendation acceptance, post-remediation gains, annotation reliability, or generalization beyond Alibaba Cloud. Those questions require longitudinal and independently reviewed studies.

\section{From Industrial Records to an Exploratory Benchmark}
\label{sec:benchmark}

The release process produces more than a gate decision. For every reviewed testcase, it retains the complete operational-obligation inventory and the expert-confirmed covered/uncovered status of each obligation. These records make it possible to study automatic assessors that reproduce the end-to-end judgment established by the reviewed process. We package a subset of them as \emph{SkillAdeqBench} and release the dataset and evaluator at \url{https://github.com/dawnvince/SkillAdqBench}.

\subsection{Benchmark Composition}

SkillAdeqBench contains 32 workflow-oriented Alibaba Cloud Skills and 123 normalized testcases. Its split is disjoint by Skill. The training set contains 16 Skills, 64 testcases, and 872 reviewed operation-status records; the test set contains 16 different Skills, 59 testcases, and 840 records. A testcase references its complete Skill package and contains a prompt, an initial \texttt{resource\_state}, and \texttt{expected\_decisions} that specify the user responses supplied when the Skill requests a choice, confirmation, authorization, or other required input. Its reference output contains the complete operation inventory for the Skill, with each operation represented by \texttt{command}, \texttt{label}, and \texttt{result}, and assigned a \texttt{cover\_status} of \texttt{covered} or \texttt{uncovered} for that scenario.

\subsection{End-to-End Assessment Task}

Given a Skill package $S$ and a testcase $t=(p,r,d)$, an assessor must produce a set
\[
  \widehat{Y}(S,t)=
  \{\langle \widehat{\alpha},\widehat{\ell},\widehat{q},\widehat{z}\rangle\},
  \qquad \widehat{z}\in\{\mathsf{covered},\mathsf{uncovered}\},
\]
where \(\widehat{\alpha}\), \(\widehat{\ell}\), and \(\widehat{q}\) are serialized as \texttt{command}, \texttt{label}, and \texttt{result}. The output must contain the complete operation inventory inferred from $S$, including operations not exercised by $t$, and assign exactly one status to every predicted operation. The assessor is not given a reference operation list; recovering the assessment units and determining their scenario-conditioned statuses form one end-to-end task.

\begin{tcolorbox}[
  colback=gray!3,
  colframe=black!45,
  boxrule=0.5pt,
  arc=1mm,
  title={Simplified testcase example},
  fonttitle=\bfseries
]
\small
A Skill instructs the agent to check for a suitable VPC, ask whether to reuse it or create a new one, follow the selected branch, deploy an application, and validate reachability. A testcase specifies:
\begin{itemize}[leftmargin=*,nosep]
  \item \textbf{prompt:} deploy the application in the target region;
  \item \textbf{resource state:} a suitable VPC named \texttt{vpc-prod} exists; and
  \item \textbf{expected decision:} when asked, reuse \texttt{vpc-prod}.
\end{itemize}
Its reference output marks \emph{check for a suitable VPC}, \emph{select the existing VPC}, \emph{deploy the application}, and \emph{validate reachability} as \texttt{covered}. It marks \emph{select creation of a new VPC} and \emph{create the VPC} as \texttt{uncovered}. Both branches remain in the output because the benchmark evaluates whether the assessor recovers the complete Skill-level operation inventory as well as the path selected by this testcase.
\end{tcolorbox}

\subsection{Evaluator and Primary Metric}

The evaluator compares a predicted record set $P_t$ with the reference set $G_t$ for each testcase. A fixed semantic matcher first establishes a one-to-one alignment $M_t\subseteq G_t\times P_t$ according to the operation semantics expressed by \texttt{command}, \texttt{label}, and \texttt{result}; \texttt{cover\_status} is not used as a matching criterion and is compared programmatically after alignment. Let $C_t$ be the number of aligned pairs whose \texttt{cover\_status} values agree. End-to-end counts are
\[
\begin{aligned}
  \mathit{TP}_t &= C_t,\\
  \mathit{FP}_t &= (|P_t|-|M_t|)+(|M_t|-C_t),\\
  \mathit{FN}_t &= (|G_t|-|M_t|)+(|M_t|-C_t).
\end{aligned}
\]
An unmatched prediction is therefore a false positive, an unmatched reference operation is a false negative, and an aligned operation with the wrong status contributes one of each. We compute precision, recall, and $F_1$ from these joint counts for every testcase and use macro-averaged testcase $F_1$ as the primary benchmark metric; a prediction with no true positives receives $F_1=0$. This scoring rule gives credit only when an assessor identifies the same operation and assigns the same scenario-conditioned status.

The released evaluator uses a frozen, temperature-zero Qwen3.7-Max semantic matcher under a one-to-one output schema. JSON syntax is repaired when possible, and the call is retried when schema validation still fails. The matching procedure and model configuration are held fixed across evaluated assessors.

SkillAdeqBench is an exploratory artifact derived from one provider's reviewed production records. It provides a concrete starting point for developing and comparing automatic Skill Test Adequacy assessors. Its current scope is workflow-oriented Alibaba Cloud Skills; broader validation requires Skills, annotation practices, and operational environments from additional providers and domains.

\section{Related Work}

\subsection{Test Adequacy Criteria Across Software Artifacts}

Test adequacy criteria provide objective measures of test-suite scope relative to a chosen artifact or behavior model~\cite{zhu1997coverage}. Software engineering research has introduced specialized criteria when conventional code coverage does not represent the behavior of interest. Examples include simulation-based criteria for distributed systems~\cite{rutherford2006simulation}, DOM-based criteria for web applications~\cite{mirzaaghaei2014dom}, REST API coverage criteria~\cite{martinlopez2019rest}, and surprise adequacy for deep learning systems~\cite{kim2019surprise}. We extend this artifact-specific perspective to workflow-oriented Cloud Skills, whose test obligations are implicit in natural-language operational content.

\subsection{Requirements Traceability and Requirements-Based Testing}

Requirements traceability connects requirements to downstream artifacts such as code, tests, and verification results. T-BERT uses pretrained language models to recover trace links between source code and natural-language artifacts~\cite{lin2021traceability}. Requirements-based testing has also generated failure-revealing testcases for industrial drivability requirements~\cite{formica2023drivability}. Guo et al.\ use a knowledge graph and automated-testing results to generate requirements for crowdsourced Android testing~\cite{guo2020crowdsourced}. Skill Test Adequacy differs because the source package does not provide an explicit requirement list: the operational obligations and their testcase links must both be recovered and reviewed.

\subsection{Natural-Language Specification Analysis}

Recent work uses AI to analyze natural-language requirements, including question-answering assistance~\cite{ezzini2023qassist} and translation into temporal-logic specifications~\cite{ma2025req2ltl}. These studies show that natural-language software artifacts can support specialized automated analyses. Our target is test-adequacy assessment rather than requirement QA or formalization.

\subsection{Agent and Skill Evaluation}

Agent benchmarks evaluate task completion, tool use, software repair, and environment interaction. Skill evaluations similarly measure whether supplying a Skill improves agent performance. These evaluations assess the effectiveness of the agent--Skill system on sampled tasks. Skill Test Adequacy instead examines the scope of the Skill's own test suite. The two forms of evidence are complementary.

\section{Conclusion}

Production Cloud Skills are delivered and maintained as operational capabilities for AI agents, yet passing existing testcases does not reveal which specified behaviors remain untested. This paper introduced Skill Test Adequacy, operationalized it as a reviewed release practice, and applied it as a prerequisite to task-success testing and subsequent publication in the Alibaba Cloud release process. Among 157 initial adequacy assessments, 57 (36.3\%) fall below the mandatory 80\% gate and 76 (48.4\%) remain below the recommended 90\% level. The 132 recommendation reports contain 639 obligation-level recommendations. These results show that test gaps observed before remediation are operationally material and that an obligation-level report provides evidence beyond a scalar score. SkillAdeqBench makes a reviewed subset available for exploratory research on automatic adequacy assessors. Future work should reduce review effort, evaluate reviewer agreement, and measure adequacy changes after remediation.

\appendix
\section{Agent Instruction Used in the Adequacy Audit}
\label{app:agent-prompt}

The following instruction is the English version of the prompt used to guide Qwen3.7-Max agent instances during operation recovery and testcase adequacy proposal. Invocation-time messages additionally provide the concrete Skill package, testcase prompt, resource state, expected user decisions, and writable working directory. Reviewers establish the final adequacy records as described in Section~3.

\begin{agentpromptbox}[Cloud Skill Test Adequacy Assessment]
\footnotesize
Use a ReAct-style observation, analysis, and action loop to assess the test adequacy of a Cloud Skill. In each round, first observe the available Skill files or testcase information, then analyze which facts are still missing, and subsequently read the required files or inspect prior judgments. Continue until both the operation inventory and the adequacy records are stable.

\medskip
\textbf{Task.}
This task can be understood by combining requirements coverage and code coverage in software engineering:
\begin{itemize}[leftmargin=*]
  \item As in requirements coverage, first recover the complete set of testable behaviors specified by the Cloud Skill package.
  \item As in code coverage, treat each core operation as an assessment unit and determine whether the testcase scenario is expected to exercise it under the Skill instructions.
  \item The testcase prompt, \texttt{resource\_state}, and \texttt{expected\_decisions} jointly play the role of the test input and scenario specification: they select the applicable workflow and its conditional path.
  \item \texttt{cover\_status} records whether the selected scenario is expected to exercise the corresponding core operation.
\end{itemize}

Your goal is first to establish the complete core-operation inventory for the Skill and then produce a complete adequacy output that marks every operation as covered or uncovered by the current testcase.

The result must be written to \texttt{output/prediction.json} in the working directory. The task is complete only when this file has been created successfully, can be read, and satisfies the required format. Showing, describing, or pasting the result in the agent response does not replace writing the file.

The inputs include:
\begin{itemize}[leftmargin=*]
  \item a Cloud Skill package;
  \item a testcase prompt; and
  \item the testcase's pre-execution \texttt{resource\_state}; and
  \item the testcase's \texttt{expected\_decisions} at user-interaction points.
\end{itemize}

Complete two tasks:
\begin{enumerate}[leftmargin=*]
  \item identify the complete set of core operations defined by the Skill; and
  \item determine the scenario-conditioned coverage status of every core operation for the current testcase.
\end{enumerate}

A \emph{core operation} is a testable operation that directly advances cloud-task execution, changes the task path, or validates the task result. Core operations include:
\begin{itemize}[leftmargin=*]
  \item querying, creating, updating, configuring, or deleting cloud resources;
  \item executing an official script supplied by the Skill, including \texttt{.sh} and \texttt{.py} files; each script as a whole counts as one core operation;
  \item invoking a CLI, API, or external system specified by the Skill;
  \item obtaining a user choice, confirmation, authorization, or required input that affects subsequent execution; and
  \item validating an execution result or performing required recovery or cleanup.
\end{itemize}

Each core operation must be a semantic unit for which it is possible to independently determine whether the current testcase scenario is expected to exercise it under the Skill instructions.

\medskip
\textbf{Execution Procedure.}
Follow the steps below in order. Keep the reasoning process within the current agent session. The final result must be saved through a file-writing tool at the required path; do not finish the task before the file has been written successfully.

\medskip
\textbf{1. Observe and understand the Skill.}
Read all relevant files in the Skill package. Understand the tasks supported by the Skill, its operational steps, resource conditions, user choices, and expected results.

Use the current observations to compile the core-operation inventory. When the source, applicable condition, or granularity of an item is unclear, continue reading the relevant files and revise the inventory.

\medskip
\textbf{2. Observe and analyze the testcase.}
Read the testcase prompt, \texttt{resource\_state}, and \texttt{expected\_decisions}. Determine the user's goal, the initial resource state, and the user responses that select interactive branches.

The \texttt{resource\_state} describes the state before execution. It may establish whether a resource exists, whether permissions are available, or whether a configuration is already present. The \texttt{expected\_decisions} specify the responses supplied when the Skill requests a choice, confirmation, authorization, or required input.

\medskip
\textbf{3. Iteratively determine operation status.}
For every identified core operation, determine whether the current testcase scenario is expected to exercise that operation under the Skill instructions.

If a judgment depends on an unconfirmed resource condition, user choice, or Skill constraint, return to the relevant Skill file or testcase information, observe it again, and update the judgment.

Consider the following factors:
\begin{itemize}[leftmargin=*]
  \item whether the testcase prompt requests the corresponding task;
  \item whether \texttt{resource\_state} activates or blocks the condition containing the operation;
  \item whether \texttt{expected\_decisions}, permission states, resource states, or failure scenarios change the execution path; and
  \item whether the operation belongs to the workflow exercised by the current testcase.
\end{itemize}

\medskip
\textbf{4. Produce the final prediction.}
Create or overwrite exactly one output file and do not create any other output file:

\begin{center}
\texttt{output/prediction.json}
\end{center}

The file must contain a JSON array. Each element represents one core operation and must contain exactly the following fields:

\begin{quote}
\ttfamily\footnotesize
\{\\
\quad ``command'': ``string'',\\
\quad ``label'': ``string'',\\
\quad ``result'': ``string'',\\
\quad ``cover\_status'': ``covered $|$ uncovered''\\
\}
\end{quote}

The output must satisfy all of the following requirements:
\begin{itemize}[leftmargin=*]
  \item Include the complete core-operation set, not only covered operations.
  \item Set \texttt{cover\_status} to either \texttt{covered} or \texttt{uncovered}.
  \item Do not include operation identifiers, reasoning, \texttt{class}, \texttt{workflow}, \texttt{step}, or \texttt{action} fields in the final prediction.
  \item Use \texttt{command}, \texttt{label}, and \texttt{result} to express the semantics of each core operation clearly.
  \item Do not write a checklist, reasoning trace, execution trace, or any other intermediate artifact.
\end{itemize}

After writing the file, read \texttt{output/prediction.json} again and verify that the saved content is a complete and valid JSON array. Do not return the JSON only in the final response.
\end{agentpromptbox}

\bibliographystyle{ACM-Reference-Format}
\bibliography{references}

\end{document}